\documentclass[pre,aps,twocolumn,showpacs]{revtex4}
\usepackage{graphicx}

\begin{document}

\title{ Localization phenomena in a DNA double helix structure : 
A twisted ladder model}

\author{Sourav Kundu}

\email{sourav.kundu@saha.ac.in}

\affiliation{Condensed Matter Physics Division, 
Saha Institute of Nuclear Physics, 
1/AF, Bidhannagar, Kolkata 700 064, India}

\author{S. N. Karmakar}

\email{sachindranath.karmakar@saha.ac.in}

\affiliation{Condensed Matter Physics Division, 
Saha Institute of Nuclear Physics, 
1/AF, Bidhannagar, Kolkata 700 064, India}

\begin{abstract}

In this work we propose a model for DNA double helix within the 
tight-binding framework that incorporates the helicity of the 
molecules. We have studied localization properties of three DNA 
sequences,the periodic poly(dG)-poly(dC) and poly(dA)-poly(dT) 
sequences and the random ATGC sequence, all of which are coupled 
to backbone withrandom site energies representing the environmental 
fluctuations. We observe that due to helicity of DNA, electron 
transport is greatly enhanced and there exists almost a 
disorder-strength independent critical value of the hopping integral, 
that accounts for helicity of DNA, for which the electronic states 
become maximally extended. We have also investigated the effect of 
backbone energetics on the transmission and $I-V$ characteristics of DNA.

\end{abstract}


\maketitle

\section{Introduction}

In recent years interest on DNA mediated charge migration has enhanced
remarkably because its potential for the development of new generation
DNA-based nanoelectronic devices and computers. Also in biology, a precise
understanding of the mechanism of electron transport along DNA could play
an important role for the description of the processes like damage sensing,
protein binding, gene regulation, cell division, etc. However, the question
that does DNA conduct electron is quite intriguing to the physicists as well
as biologists, and, even today the electronic and transport properties of
DNA are not well understood. Inspite of the current intense debate, the 
electron transport properties of DNA were being addressed soon after the
discovery of the double helix structure of DNA by Watson and 
Crick~\cite{watson}. Eley and Spivey~\cite{eley} first suggested that DNA
could behave like an electric conductor. More recently with the advent of
measurements on single DNA molecule, Kelley {\it et al.}~\cite{kelley}
showed that DNA behaves like a conducting molecular wire. But a number of
conflicting experimental 
results~\cite{fink,porath,cai,tran,zhang,storm,yoo,guo}
and a variety of theoretical 
models~\cite{murphy,bixon,berlin,hermon,yu,cuni,roche,cuenda,peyrard,klotsa,guti2}
appeared in the literature. Experiments on periodic double stranded
poly(dG)-poly(dC) DNA sequence reported both conducting~\cite{fink} as
well as semiconducting~\cite{porath} behavior. Measurements on aperiodic
$\lambda$-phage DNA sequence suggested that it could be metallic~\cite{fink},
superconducting~\cite{kasumov} at low temperature or insulating~\cite{storm}.
Unambiguous and reproducible experimental results  are still a great
technological challenge due to the complexity of environment, thermal
vibrations, contact resistance and sequence variability of the DNA
molecules. On the other hand, lack of clear understanding of charge transfer
mechanism in DNA leads to various phenomenological models in which charge
transport is mediated by polarons~\cite{conwell}, solitons~\cite{hermon},
electrons or holes~\cite{dekker,ratner,beratan}. The electronic properties
and the conducting behavior of DNA thus remain highly controversial and
Ref.~\cite{endres} provides an excellent review on this issue.

In order to explain the diverse experimental results, one should consider 
three different contributions that are present in 
an experiment with DNA, namely, the sequence of 
the pairs of nucleotides({\it i.e.}, 
base-pairs) in DNA, the presence of 
backbones and the influence of environment. The environment in turn pieces 
out into three main parts, the substrate on which experiments are performed, 
the surrounding temperature~\cite{guti1} and  the humidity. 
As the sugar-phosphate backbones are negatively charged and hanging outside 
the double-helix structure of DNA, they can easily interact 
with the substrate and the effective on-site energies of the backbone get 
modulated to a great extent~\cite{tran,zhang,storm,kasumov,lcai,pablo}. 
The experiments with DNA are performed mostly in two different conditions: 
{\it i}) in natural aqueous buffer solutions, and {\it ii}) in dry conditions.  
In DNA's natural aqueous buffer solution, the backbone phosphates usually 
attract counter-ions and polar water molecules to neutralize the phosphates, 
and in the process modify their ionization potential~\cite{barnett}. Even 
for the dry case in vacuum, there are few counter-ions or water molecules 
that may still reside on the phosphates of the backbones even after 
the drying process and hence can change the effective on-site energy of
the backbone sites. Though the main conducting pathway is believed to be
the $\pi -\pi$ interaction of the stacked base-pairs in double stranded 
DNA and the backbones do not take part directly in the electronic 
transport process of DNA, one can significantly control the transport 
behavior of DNA just by tuning the environment which effectively modifies
the backbone on-site energies introducing disorder. We have shown that this
kind of control can ignite a semiconducting to metal transition in the
conducting behavior of DNA.

In this paper we study the electronic conduction properties of DNA
double helix within the tight-binding framework where environmental
fluctuations are modelled in terms of disordered on-site potentials 
of the backbone sites. We propose a model that explicitly takes into
account the helicity of the DNA molecule. We have observed that 
helicity significantly enhances electronic conduction, and, the 
interplayof helicity and backbone disorder has nontrivial effects 
on the transmission characteristics of the DNA molecules.

This paper is organized as follows. In Sec. II we introduce the model 
Hamiltonian and briefly describe our theoretical formulation. We analyze
our numerical results in Sec. III and finally conclude in Sec. IV.

\section{Model and Theoretical Formulation}

\begin{figure}[ht]
\centering

    \includegraphics[width=30mm,height=80mm]{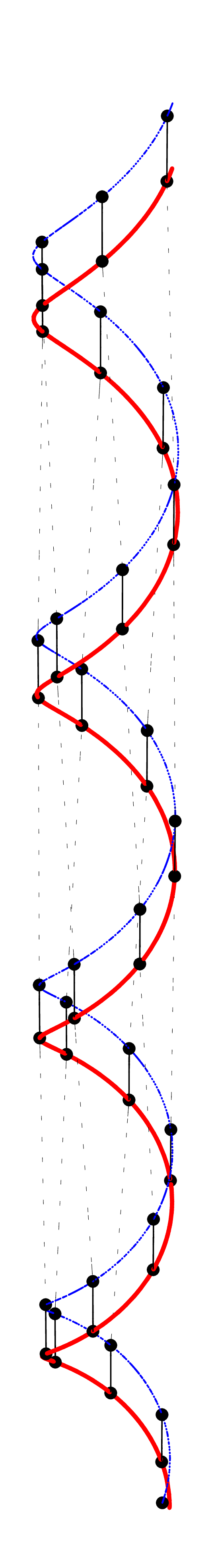}

\caption{(Color online). Schematic view of the twisted ladder model 
for electronic transport along ds-DNA. The red (thick) and 
blue (thin) lines are the two strands and the solid dots on them 
are the nucleotides. The solid lines between the solid dots represent 
the interstrand hopping ($v$) between neighboring nucleotides within 
a given pitch and dashed lines represent the interstrand 
hopping ($v'$) between neighboring nucleobases of adjacent pitches. 
Backbones are not shown in the figure.}

\label{fig1}

\end{figure}

DNA, carrier of genetic code of all living organism, is a long 
and complex biological macro-molecule, a $\pi$-stacked array of
base-pairs made from four nucleotides guanine (G), adenine (A), 
cytosine (C)and thymine (T)  coupled via hydrogen bond and forms 
a double-helix structure associated with sugar-phosphate backbone 
attached to each base-pair~\cite{klotsa}. In most of the theoretical 
models it is assumed that electronic transport~\cite{cuni,zhong,bakhshi,ladik} 
is through the long-axis of the DNA molecule. In the present study, 
the helicity of the DNA molecule is incorporated in a 
tight-binding (TB) dangling backbone ladder model~\cite{klotsa,gcuni} 
by adding hopping integrals due to the proximity of atoms 
in the upper strand with the corresponding atoms of the 
lower strand in the next pitch (see Fig.~\ref{fig1}).  
The Hamiltonian for the twisted ladder model can be expressed as 

\begin{equation}
 H_{DNA}=H_{ladder}+ H_{helicity}+H_{backbone}~,
\label{hamilton}
\end{equation}

\noindent
where, 
\begin{eqnarray}
& H_{ladder}&= \sum\limits_{i=1}^N\sum\limits_{j=I,II}\left(\epsilon_{ij}
c^\dagger_{ij}c_{ij}
+t_{ij}c^\dagger_{ij}c_{i+1j}+\mbox{H.c.} \right)\nonumber \\
&&~~~~~~~~~~~~~~~~~+ \sum_{i=1}^N v \left(c^\dagger_{iI}c_{i II}+
\mbox{H.c.} \right)~, \\
& H_{helicity}&=\sum\limits_{i=1}^N v^{\prime} 
\left(c^\dagger_{i II}c_{i+n I}+ \mbox{H.c.} \right)~, \\
& H_{backbone}&=\sum\limits_{i=1}^N\sum\limits_{j=I,II}
\left(\epsilon_i^{q(j)}c^\dagger_{i q(j)}c_{i q(j)}\right.\nonumber \\
&&~~~~~~~~~~~~~~~~~+\left.t_i^{q(j)}c^\dagger_{ij}c_{i q(j)}+
\mbox{H.c.} \right)~,
\end{eqnarray} 
where $c_i^\dagger$ and $c_i$ are the creation and annihilation 
operators for electrons in the {\it i}th Wannier state, 
$t_{ij}=$ hopping integral between neighboring nucleotides along each 
strand of the ladder, $\epsilon_{ij}=$ on-site potential energy of the 
nucleotides, $t_{i}^{q(j)}=$ hopping amplitude between a nucleotide 
and the corresponding backbone site, $\epsilon_{i}^{q(j)}=$ on-site 
potential energy of the backbone sites with $q(j)=\uparrow,\downarrow$ 
for $j$=I and II respectively denoting upper and lower backbone sites, 
$v=$ interstrand hopping integral between two neighboring sites  of DNA 
within a given pitch, $v'=$ interstrand hopping integral between 
neighboring atomic sites in the adjacent pitches which actually 
accounts for the helical structure of DNA. Here $n$ is the number 
of sites in each strand within a given pitch.
For simplicity we set $\epsilon_i^{q(j)}=\epsilon_b$, $t_{ij}=t_i$ 
and $t_i^{q(j)}=t_b$.           

         It is possible to obtain an analytical expression for the
dispersion relation of an infinite homogeneous DNA chain (setting
$\epsilon_{ij}=\epsilon$ and $t_i=t$)  
modeled by twisted ladder in the absence backbones. 
Using Bloch's theorem, the dispersion relations for the highest occupied 
molecular orbital (HOMO)($E_{-}$) and the
lowest unoccupied molecular orbital (LUMO)($E_{+}$) can be expressed as 
$E_{\pm}=\epsilon+2t\cos (k)\pm\sqrt{v^2+v'^2+2vv'\cos (nk)}$, 
which are separated by an energy gap $E_g=2 \sqrt{v^2+v'^2+2vv'\cos (nk)}$. 
Notice that for fixed $v$ and $v'$, $E_g$ explicitly depends on $n$.

         In order to study the transport behavior of DNA, 
we use semi-infinite 1D chains as leads connected cross-wise to 
the left (L) and right (R) ends of the DNA double helix and 
the Hamiltonian of the entire 
system is given by 
$ H=H_{DNA}+H_L+ H_R + H_{tun}$. 
The explicit form of $H_L$, $H_R$  and 
$H_{tun}$ are 
\begin{eqnarray}
& H_L & =\sum\limits_{i=-\infty}^0\left(\epsilon c^\dagger_ic_i+
t c^\dagger_{i+1}c_i+\mbox{H.c.} \right)~, \\
& H_R & =\sum\limits_{i=N+1}^\infty\left(\epsilon c^\dagger_ic_i+
t c^\dagger_{i+1}c_i+\mbox{H.c.} \right)~, \\
& H_{tun}& = \tau \left(c^\dagger_0c_1+c^\dagger_Nc_{N+1} +
\mbox{H.c.}\right)~,
\end{eqnarray}
where $\tau$ is the tunneling matrix element between DNA and the leads.         

	We use the Green's function formalism in order to obtain transmission 
probability $T(E)$ of electron~\cite{datta1,datta2}
through DNA. The single particle
retarded Green's function operator for the entire system at energy $E$ 
is given by $G^r=(E-H+i\eta)^{-1}$, where $\eta\rightarrow0^+$. 
The transmission probability is given by 
$T(E)={\mbox {\rm Tr}} [\Gamma_L G^r \Gamma_R G^a]$, 
$E$ being the incident 
electron energy and the trace is over the reduced Hilbert space spanned by the 
DNA molecule. The retarded and the advanced Green's functions in the
reduced Hilbert space can be expressed as 
$G^r=[G^a]^\dagger=[E- H_{DNA}-\Sigma^r_L-\Sigma^r_R+i\eta]^{-1}$, where 
$\Sigma^{r(a)}_{L(R)}=H^\dagger_{\mbox{tun}} G^{r(a)}_{L(R)} 
 H_{\mbox{tun}}$ 
and $\Gamma_{L(R)}=i[\Sigma^r_{L(R)}-\Sigma^a_{L(R)}]$, 
$G^{r(a)}_{L(R)}$ 
being the retarded(advanced) Green's function for the left(right) lead. 
Here $\Sigma^r_{L(R)}$ and $\Sigma^a_{L(R)}$ are the retarded and advanced 
self-energies of the left(right) lead  due to its coupling with the DNA
molecule. It can be easily shown that 
$\Gamma_{L(R)}=-2~{\mbox{\rm Im}} (\Sigma^r_{L(R)})$.             

         Restricting ourselves within the linear response regime, at 
absolute zero temperature, the conductance $g$ is given by the two-terminal
Landauer formula $g=\frac{2e^2}{h}T(E_F)$ and the current for an applied
bias voltage $V$ is given by 
\begin{equation}
I(V)=\frac{2e}{h} \int^{E_F+eV/2}_{E_F-eV/2} T(E)dE~, 
\end{equation}
$E_F$ being the Fermi energy. Here we have assumed that there is no charge 
accumulation within the system, {\it i.e.}, the voltage drop occurs only 
at the boundaries of the conductor.

\section{Results and Discussions}

     We first study the localization properties of the system. The 
localization length $l$ of the system is calculated from the 
Lyapunov exponent~\cite{ventra}
\begin{equation}
\gamma = 1/l = -\lim\limits_{\Lambda\to\infty}\frac{1}{\Lambda}<\ln(T(E))>~,
\end{equation}
where $\Lambda$ = length of the system in terms of basepairs,  
and $ <> $ denotes average over different disorder configurations. In actual 
experimental situations there are various environmental fluctuations. 
We have simulated these environmental fluctuations in the model by 
considering the backbone site energy $\epsilon_b$ to be randomly 
distributed within the range [$\bar\epsilon_b$-w/2, $\bar\epsilon_b$+w/2],
where $\bar\epsilon_b$ is the average backbone site energy and
w represents the disorder strength. For the purpose of numerical 
investigation the on-site energies of the nucleotides are taken as the 
ionization potentials, and the following numerical values are used through 
out this work: $\epsilon_G=-0.56 eV$, $\epsilon_A=-0.07 eV$, 
$\epsilon_C= 0.56 eV$, $\epsilon_T= 0.83 eV$. The intrastrand hopping 
integrals between like nucleotides are taken as $t=0.35eV$ while those between 
unlike nucleotides are taken as $t=0.17 eV$. We take interstrand hopping 
parameter to be $v=0.3 eV$. As all the nucleobases are connected 
with sugar-phosphate backbone by identical C-N bonds, 
the corresponding hopping parameter between the nucleobase and backbone is 
taken to be equal for all the cases and we take $t_b=0.7 eV$~\cite{cuenda}. 
The parameters used here are the same as those used in~\cite{guo} which
were  extracted from the {\it ab initio} calculations~\cite{voit,yan,senth}.

\begin{figure}[ht]

  \centering

  \begin{tabular}{cc}

    \includegraphics[width=40mm,height=30mm]{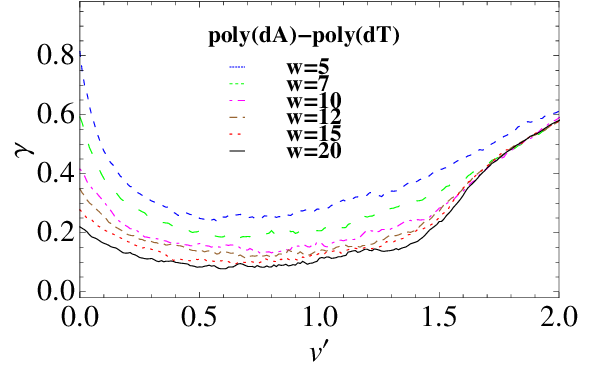}&
   
    \includegraphics[width=40mm,height=30mm]{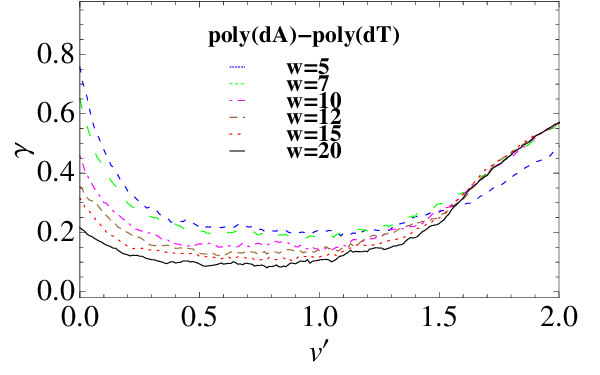}\\

    \includegraphics[width=40mm,height=30mm]{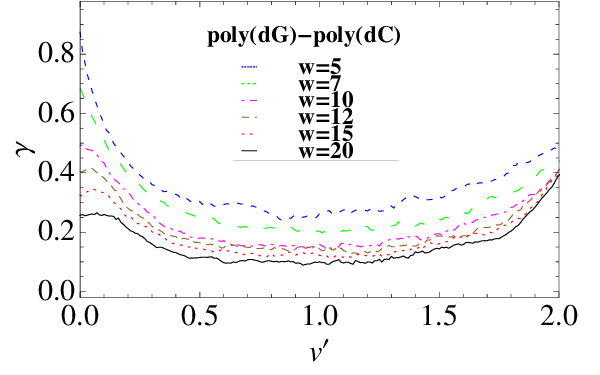}&

    \includegraphics[width=40mm,height=30mm]{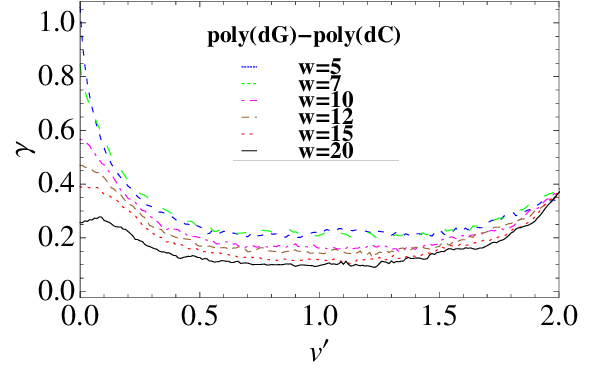}\\

    \includegraphics[width=40mm,height=30mm]{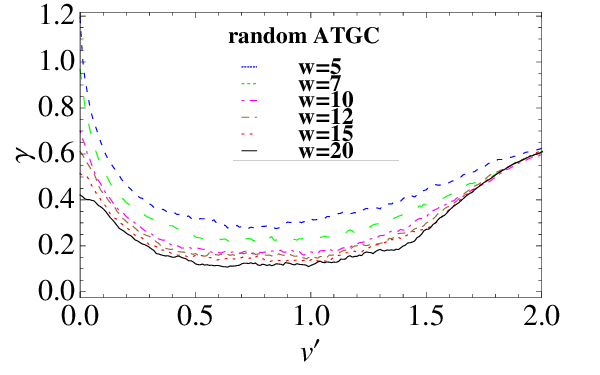}&

    \includegraphics[width=40mm,height=30mm]{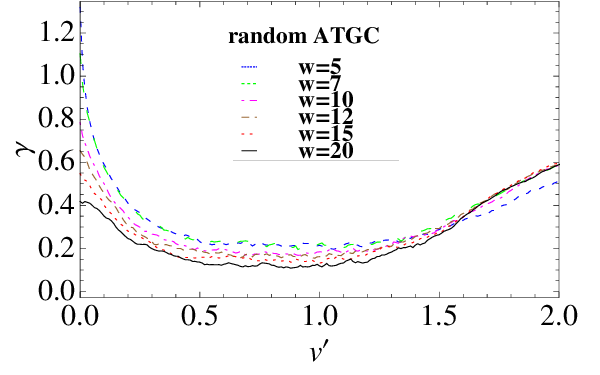}

 \end{tabular}
\caption{(Color online). Lyapunov exponent $\gamma$ vs $v'$ for 
three DNA sequences at six different disorder strengths (w).
The left column shows plots for $\bar\epsilon_b$ = 0 eV and 
the right column for $\bar\epsilon_b$ = 3 eV. The critical 
values of $v'$ (say, $v'_c$)  corresponding to the minima 
of $\gamma$ are almost insensitive to the disorder strength.}

\label{fig2}

\end{figure}

    In Fig.~\ref{fig2} we have plotted inverse localization 
length $\gamma$ of the periodic poly(dG)-poly(dC) and 
poly(dA)-poly(dT) sequences and the random ATGC sequence 
with respect to $v'$ (where $v'$ accounts for the helicity of DNA) 
for various values of the  disorder strength w. 
It has been observed that all the curves have a general shape for the 
periodic as well as random DNA sequences. The variation 
of $\gamma$ with $v'$ is not monotonic, there exists a flat 
minimum of $\gamma$ for each sequence and the position of this 
minimum for a specific sequence is almost  insensitive to the disorder 
strength w. As we increase $v'$ starting from zero, $\gamma$ starts 
decreasing, which is quite natural because by allowing $v'$ 
we are opening new channels for electron conduction, so the 
system becomes much more conductive and $\gamma$ decreases. 
Now there is a flat minimum which signifies that in this region 
the system is most conducting as both the channels, one along 
the main DNA ladder and one arising due to helicity,are 
contributing almost equally. 

Then $\gamma$ starts increasing implying that $v'$ channel starts to 
dominate over the other channels due to $t$ and $v$, and as we further 
increase $v'$ the system becomes more and more like a 1D disordered one 
(as the role of $t$ and $v$ becomes negligible). As in the 1D systems
Anderson localization sets in if we introduce small amount of disorder 
into the system, the same thing is also happening here as we 
increase $v'$ beyond the minimum, $\gamma$ starts increasing 
indicating that the system effectively becoming a 1D disordered one.
One may ask why these minima of $\gamma$ versus $v'$ plots are almost 
insensitive to disorder strength. The minimum of any $\gamma$ versus $v'$ 
curve is a result of the competition of two channels, the $v'$ channel and 
resultant of $t$ and $v$ channels. Disorder can change the magnitude 
of $\gamma$ but it has nothing to do with the nature of this competition, 
so the minima occurs almost at the same value of $v'$ irrespective of the
disorder strength.    

\begin{figure}[ht]
\centering

  \begin{tabular}{c}

    \includegraphics[width=60mm]{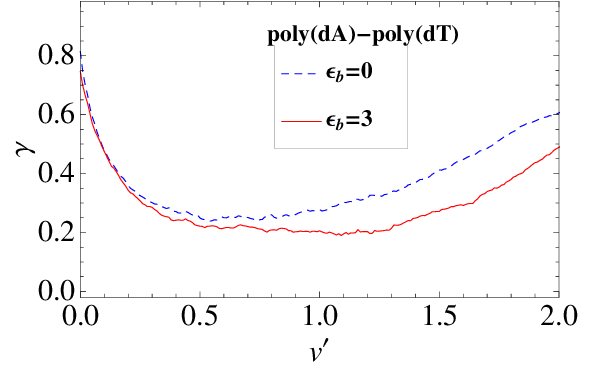}\\
   
    \includegraphics[width=60mm]{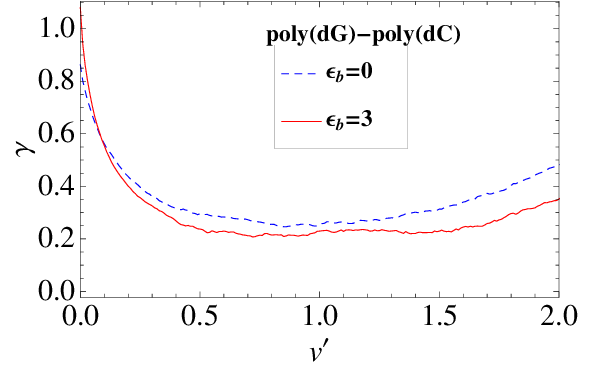}\\

    \includegraphics[width=60mm]{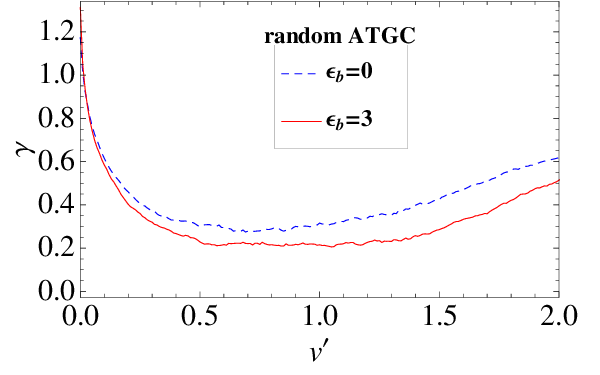}

  \end{tabular}

\caption{(Color online). Lyapunov exponent $\gamma$ vs $v'$ for three 
DNA sequences at a fixed disorder strength (w=5) for
two cases $\bar\epsilon_b$ = 0 eV and $\bar\epsilon_b$ = 3 eV. 
Figure shows that there is a shift in positions of the 
minima for $\gamma$ vs $v'$ curves depending on the value 
of the average backbone site-energy ($\bar \epsilon_b$).}

\label{fig3}

\end{figure}

     In Fig.~\ref{fig3}
we have plotted $\gamma$ versus $v'$ for a fixed disorder
strength (w=5) for various values of $\bar\epsilon_b$. 
It is to be noted that here the minima are sensitive to 
the average backbone site energy($\bar\epsilon_b$), 
the minima get shifted as we increase $\bar\epsilon_b$. 
To explain this one should look at the renormalized 
site energies of the basepairs, as we decimate the backbone sites. 
Let us look at the expression for the renormalized site energy 
of the nucleobases $\epsilon^\prime_i=\epsilon_i-\frac {t_b^2}{\epsilon_b-E}$, 
the fluctuation in $\epsilon_i$ due to disorder in backbone site energy
becomes smaller as we increase $\bar\epsilon_b$ at a given w. 
Thus if we consider two cases $\bar\epsilon_b$=0 and 3, we have less 
scattering for $\bar\epsilon_b=3$ case, and accordingly, $v'$ 
corresponding to its minimum has to be grater than that of the 
$\bar\epsilon_b=0$ case to dominate over the other channels 
and hence the minimum for $\bar \epsilon_b$=3 getsshifted. 

\begin{figure}[ht]

  \centering

  \begin{tabular}{c}

    \includegraphics[width=60mm]{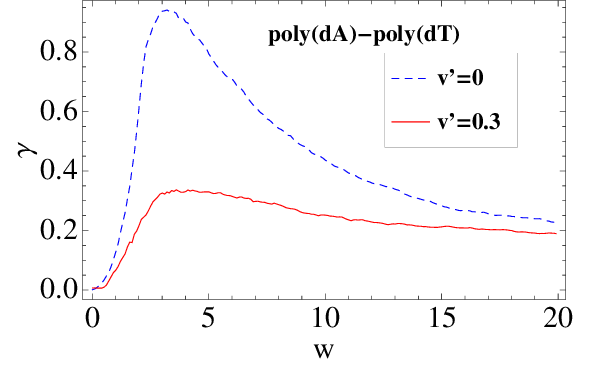}\\
   
    \includegraphics[width=60mm]{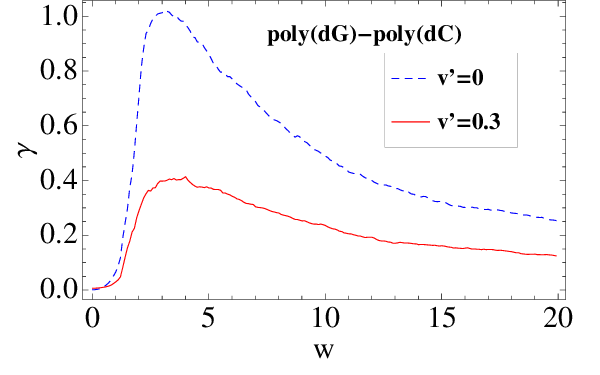}\\

    \includegraphics[width=60mm]{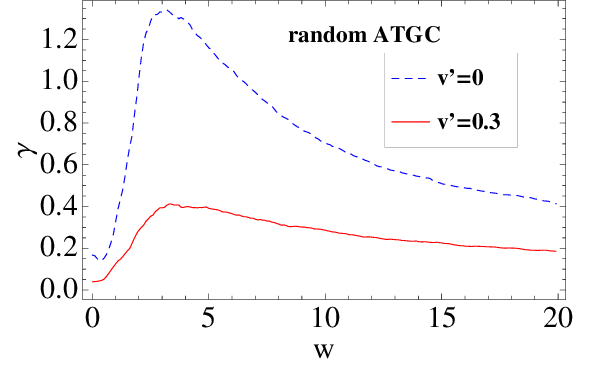}\\

  \end{tabular}

\caption{(Color online). Lyapunov exponent $\gamma$ vs 
w for the three DNA sequences for the two cases $v'=$0 eV and 
$v'=$0.3 eV showing significant drop in $\gamma$ for higher $v'$.}

\label{fig4}

\end{figure}
      In Fig.~\ref{fig4} we have shown the behavior of $\gamma$ 
with disorder strength $w$ for two cases with $v'=0$ and $v'=0.3$. 
It is clearly visible that for the higher value of $v'$,
there is a significant drop in $\gamma$ which is obvious because by 
introducing $v'$ we are adding other channels for 
conduction and  the localization length of the system increases. 
Initially if the localization length for $v'=0$ was a few base-pairs in 
presence of backbone disorder, it increases three to four times 
if we consider the effect of helicity of the DNA molecules.
There is another thing to be noted is that there exists a maximum in each
of these curves, beyond which $\gamma$ starts decreasing even if we
increase $w$. This feature was already reported by Guo {\it et al.}
for the fishbone and dangling backbone ladder models~\cite{guo} 
and the same effect is also present in our model, but this anomalous 
effect of backbone disorder is much smaller in our model which inturn 
also signify that helical nature of DNA can minimize the effect of 
disorder {\it i.e.,} environmental fluctuations.

\begin{figure}[ht]

  \centering

  \begin{tabular}{c}

    \includegraphics[width=61mm]{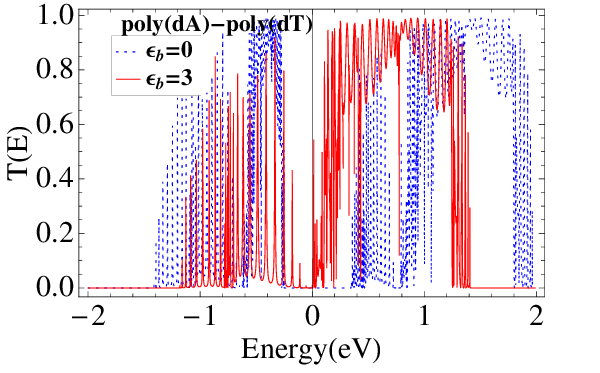}\\
   
    \includegraphics[width=61mm]{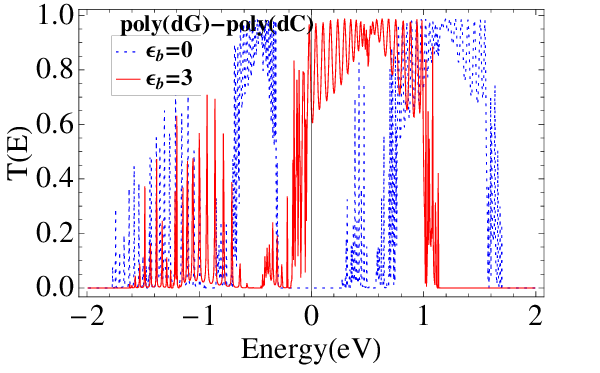}\\

  \end{tabular}
\caption{(Color online). Transmission coefficient as a function 
of energy for the poly(dA)-poly(dT) and poly(dG)-poly(dC) sequences
for the cases $\bar\epsilon_b$ = 0 eV and $\bar\epsilon_b$ = 3 eV 
without backbone disorder ({\it i.e.}, w=0) with $v'$ = 0.3 eV.}

\label{fig5}

\end{figure}

\begin{figure}[ht]

  \centering

 \begin{tabular}{c}

    \includegraphics[width=61mm]{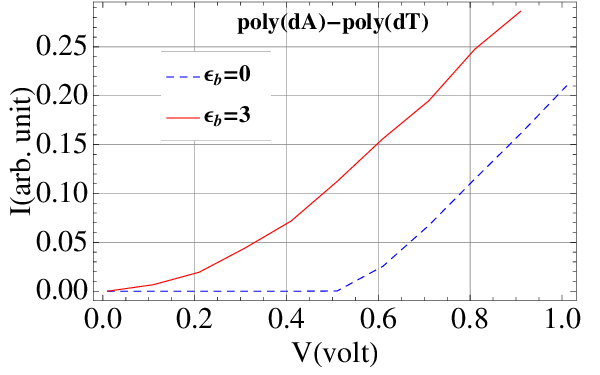}\\
   
    \includegraphics[width=61mm]{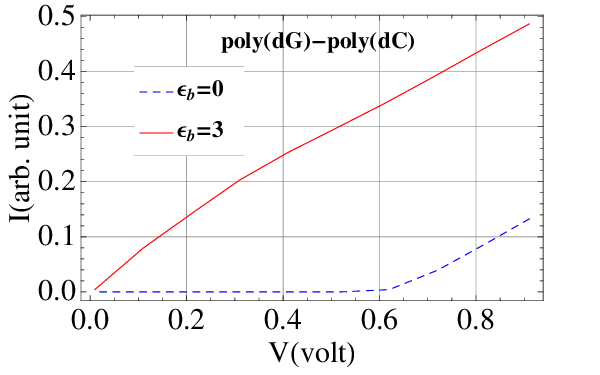}\\

  \end{tabular}

\caption{(Color online). $I-V$ characteristics for the 
poly(dA)-poly(dT) and poly(dG)-poly(dC) sequences for the 
cases $\bar\epsilon_b$ = 0 eV and $\bar\epsilon_b$ = 3 eV without 
backbone disorder ({\it i.e.}, w=0) with $v'$ = 0.3 eV.}

\label{fig6}

\end{figure}
\begin{figure}[ht]

  \centering

 \begin{tabular}{c}

    \includegraphics[width=60mm]{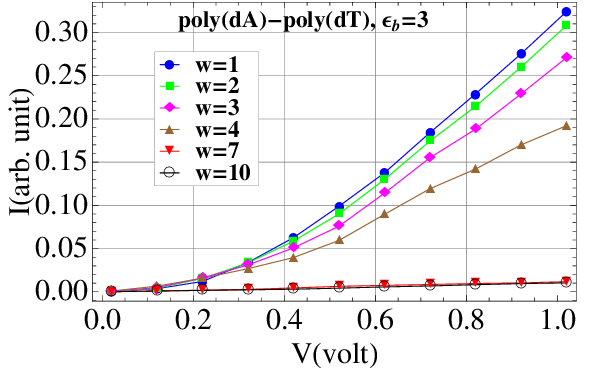}\\
   
    \includegraphics[width=60mm]{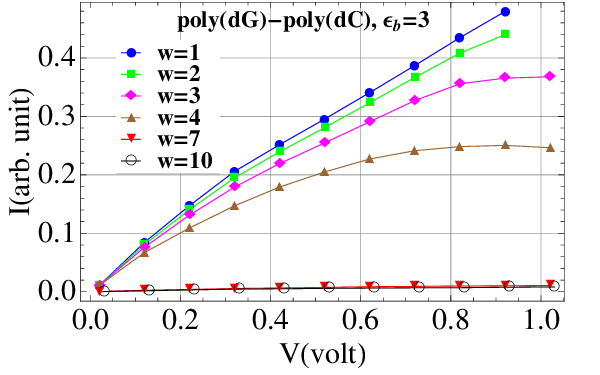}\\

  \end{tabular}

\caption{(Color online). $I-V$ characteristics for the  
poly(dA)-poly(dT) and poly(dG)-poly(dC) sequences for 
the $\bar\epsilon_b$ = 3 eV case in presence of backbone 
disorder (w) with $v'$ = 0.3 eV.}

\label{fig7}

\end{figure}

   We have also investigated the $I-V$ response of the two periodic sequences. 
The temperature is set to 0 K. To minimize the contact effects we choose 
tunnelling parameter $\tau$ to be optimum $\it{i.e.}$, $\tau$=$\sqrt{t_{ij}\times t}$ 
between ds-DNA and the electrodes, where t is the hopping parameter 
for the electrodes~\cite {macia}. In Fig.\ref{fig5} 
we have shown variation of transmission probability $T$ with respect 
to  energy $E$ for these two sequences, and it clearly shows that 
by tuning the backbone site energy one can control the energy gap 
of the system. In Fig.~\ref{fig6} the $I-V$ characteristics are 
shown without backbone disorder and it has been observed that for 
$\bar\epsilon_b=3 $, the cut-off voltage becomes nearly equal to 
zero for both the periodic poly(dG)-poly(dC) and poly(dA)-poly(dT)
sequences. For poly(dG)-poly(dC) case the response is almost linear which
indicates that the system would undergo a transition from 
semiconducting to metallic phase depending on on-site energies of the 
backbone, not only disorder can induce that kind of transition~\cite{guo}. 
$I-V$ characteristics for $\bar\epsilon_b=0$ case are in good agreement with 
experimental results~\cite{porath}. We have checked the $I-V$ response 
for different system sizes, and observed that there is no significant 
change in the characteristics. In Fig.\ref{fig7} we show the $I-V$ responses 
for the two periodic sequences for $\bar\epsilon_b=3$ case in presence of 
the backbone disorder. It clearly shows that with increasing disorder 
the current in the system decreases.

\section{Concluding Remarks}
	Within the tight-binding framework, the
fishbone and the dangling backbone ladder models are 
generally used to study the transport properties of DNA like 
systems, but none of these models have incorporated the 
effect of helicity of DNA molecules. Helicity is a very fundamental 
aspect of the DNA structure and gives the possibility of new conduction
channels in DNA. In this paper we propose a model that accounts 
for the helical structure of the DNA molecules and show that
there is a significant change in the localization properties 
and $I-V$ response of the systems. One of the main results is
that there exists a almost disorder independent critical 
hopping ($v'_{c}$) for which localization length becomes 
maximum for each of the sequences that we have considered. 
At this critical hopping strength system is least affected by the external 
disturbances, {\it i.e.}, environmental effects are least at this point.
If one can utilize this information properly then it might be possible to  
minimize the environmental effects in the actual experiments. We have also
shown that though backbones do not play any role directly
in electron conduction but they can significantly contribute by narrowing 
the energy gap and reducing the cut-off voltage. It might also be possible 
to find DNA in a complete metallic phase even in presence of environmental
fluctuations as evident from the almost linear 
$I-V$ response of poly(dG)-poly(dC) sequence for $\bar\epsilon_b=3$. 
We look forward that there might be experimental investigations 
as well as {\it ab initio} calculations in near future to 
find the exact value of the interstrand hopping integral ($v'$) 
between nucleobases of adjacent pitches and its actual 
role in transport.


\newpage
\begin{thebibliography}{99}

\bibitem{watson} J. Watson and F. Crick, Nature (London) \textbf{171}, 
737 (1953).

\bibitem{eley} D. D. Eley and D. I. Spivey, Trans. Faraday Soc. \textbf{58},
411 (1962).

\bibitem{kelley} S. O. Kelley and J. K. Barton, Science \textbf{283}, 
375 (1999).

\bibitem{fink} H. W. Fink and C. Sch\"{o}nenberger, Nature (London)
\textbf{398}, 407 (1999).

\bibitem{porath} D. Porath, S. De Vries, and C. Decker, Nature (London)
\textbf{403}, 635 (2000).

\bibitem{cai} L. Cai, H. Tabata, and T. Kawai, Appl. Phys. Lett.
\textbf{77}, 3105 (2000).

\bibitem{tran} P. Tran, B. Alavi, and G. Gr\"{u}ner, Phys. Rev. Lett.
\textbf{85}, 1564 (2000).

\bibitem{zhang} Y. Zhang, R. H. Austin, J. Kraeft, 
E. C. Cox, and N. P. Ong, 
Phys. Rev. Lett. \textbf{89},
198102 (2002).

\bibitem{storm} A. J. Storm {\it et al.}, Appl. Phys. Lett. \textbf{79},
3881 (2001).

\bibitem{yoo} K. H. Yoo {\it et al.}, Phys. Rev.Lett. \textbf{87},
198102 (2001).
 
\bibitem{guo} A-M Guo, S-J Xiong, Z. Yang, and H-J Zhu, 
Phys. Rev. E \textbf{78}, 061922 (2008).  

\bibitem{murphy}  C. J. Murphy, M. A. Arkin, Y. Jenkins, 
N. D. Ghatlia, S. Bossman, N. J. Turro, and J. K. Barton,  
Science
\textbf{262}, 1025 (1993).

\bibitem{bixon} M. Bixon, B. Giese, S. Wessely, T. Langenbacher, 
M. E. Michel-Beyerle, and J. Jortner, 
Proc. Natl. Acad. Sci. USA
\textbf{96}, 11713 (1999).

\bibitem{berlin} Y. A. Berlin, A. L. Burin, and M A. Ratner, J. Am. Chem.
Soc. \textbf{123}, 260 (2001).

\bibitem{hermon} Z. Hermon, S. Caspi, and E. Ben-Jacob, Europhys. Lett.
\textbf{43}, 482 (1998).

\bibitem{yu} Z. G. Yu and X. Song, Phys. Rev. Lett. \textbf{86}, 
6018 (2001).

\bibitem{cuni} G. Cuniberti, L. Craco, D. Porath, and C. Dekker, 
Phys. Rev. B \textbf{65}, 
241314(R) (2002).

\bibitem{roche} S. Roche, Phys. Rev. Lett. \textbf{91}, 108101 (2003).

\bibitem{cuenda} S. Cuenda and A. Sanchez, 
Fluctuations and Noise Letters 
\textbf{4}, L491-L504 (2004)

\bibitem{peyrard} M. Peyrard, 
Nonlinearity \textbf{17}, R1-R40 (2004).

\bibitem{klotsa} D. Klotsa, R. A. R\"{o}mer, and M. S. Turner, Biophysical
Journal \textbf{89}, 2187 (2005). 

\bibitem{guti2} R. Guti\'{e}rrez, S. Mohapatra, H. Cohen, D. Porath and G. Cuniberti, 
Phys. Rev. B
\textbf{74}, 235105 (2006)

\bibitem{kasumov} A. Y. Kasumov {\it et al.}, Science \textbf{291},
280 (2001).

\bibitem{conwell} E. M. Conwell and S. V. Rakhmanova, Proc. Natl. Acad. 
Sci. USA \textbf{97}, 4557 (2000).

\bibitem{dekker} C. Dekker and M. A. Ratner, Physics World \textbf{14}(8):
29-33 (2001).

\bibitem{ratner} M. A. Ratner, Nature (London) \textbf{397}, 480 (1999).

\bibitem{beratan} D. N. Beratan, S. Priyadarshy, and S. M Risser, Chem.
Biol. \textbf{4}, 3 (1997).

\bibitem{endres} R. G. Endres, D. L. Cox, and R. R. P. Singh, Rev. Mod. Phys.
\textbf{76}, 195 (2004).

\bibitem{guti1} R. Guti\'{e}rrez, S. Mandal and G. Cuniberti, 
Nano Lett. 
\textbf{5}, 1093 (2005)

\bibitem{pablo} P. J. de Pablo, F. Moreno-Herrero, J. Colchero, 
J. G\'{o}mez Herrero, 
P. Herrero, A. M. Bar\'{o}, P. Ordej\'{o}n, J. M. Soler, and E. Artacho, 
Phys. Rev. Lett.
\textbf{85}, 4992 (2000).  

\bibitem{lcai} L. Cai, H. Tabata, and T. Kawai, 
Nanotechnology \textbf{12}, 211 (2001).

\bibitem{barnett} R. N. Barnett, C. L. Cleveland, U. Landman, E. Boone, 
S. Kanvah, and G. B. Schuster, J. Phys. Chem. A
\textbf{107}, 3525 (2003).

\bibitem{zhong} J. Zhong, 
in {\it Proceedings of the 2003 Nanotechnology Conference}, Vol. 2. 
Edited by M. Laudon and B. Romamowicz. 
Computational Publications, CAMBRIDGE, MA. 
Nanotech 
\textbf 105-108 (2003).

\bibitem{bakhshi} A. K. Bakhshi, P. Otto, J. Ladik and M. Seel, 
Chem. Phys.
\textbf{108}, 215 (1986).

\bibitem{ladik} J. Ladik, M. Seel, P. Otto, and A. K. Bakhshi, 
Chem. Phys.
\textbf{108}, 203 (1986).

\bibitem{gcuni} G. Cuniberti, E. Maci\'{a}, A. Rodriguez, and R. A. R\"{o}mer, 
in {\it Charge Migration in DNA: 
Perspectives from Physics, Chemistry and Biology},
edited by T. Chakraborty, Springer-Verlag, Berlin
\textbf (2007).

\bibitem{datta1} S. Datta, {\it Electronic transport in mesoscopic systems}, 
Cambridge University Press, Cambridge (1995).

\bibitem{datta2} S. Datta, {\it Quantum Transport: Atom to Transistor}, 
Cambridge University Press, Cambridge (2005).

\bibitem{ventra} M. D. Ventra, {\it Electrical transport in nanoscale system},
Cambridge University Press, Cambridge (2008).

\bibitem{voit} A. A. Voityuk, J. Jortner, M. Bixon, and N. R\"{o}sch, 
J. Chem. Phys. 
\textbf {114}, 5614 (2001).

\bibitem{yan} Y. J. Yan and H. Y. Zhang, 
J. Theor. Comput. Chem. 
\textbf{1}, 225 (2002).

\bibitem{senth} K. Senthilkumar, F. C. Grozema, C. F. Guerra, F. M. 
Bickelhaupt, F. D. Lewis, Y. A. Berlin, M. A. Ratner, and L. D. A. Siebbeles, 
J. Am. Chem. Soc.
\textbf{127}, 14894 (2005).

\bibitem{macia} E. Maci\'{a}, F. Triozon, and S. Roche, Phys. Rev. B
\textbf{71}, 113106 (2005)

\end{thebibliography}
\end{document}